\documentclass[12pt]{iopart}

\usepackage{graphicx}
 \usepackage{subfig}

\begin{document}

\title[Multi-focal spherical media and geodesic lenses in geometrical optics]{Multi-focal spherical media and geodesic lenses in geometrical optics}

\author{Martin \v{S}arbort$^1$ and Tom\'{a}\v{s} Tyc$^{1,2}$}

\address{$^1$ Institute of Theoretical Physics and Astrophysics, Masaryk University, Kotl\'{a}\v{r}sk\'{a} 2, 61137 Brno, Czech Republic}
\address{$^2$ Faculty of Informatics, Masaryk University, Botanick\'{a} 68a, 60200 Brno, Czech Republic}

\ead{martin@sarbort.cz and tomtyc@physics.muni.cz}  
\begin{abstract}
This paper presents a general approach to designing the isotropic spherical media with complex spatial structure that provide different types of imaging for different light rays. It is based on equivalence of the spherical medium and the corresponding geodesic lens. We use this approach to design multi-focal gradient-index lenses embedded into an optically homogeneous region and multi-focal absolute instruments that provide perfect imaging of three-dimensional domains. 
\end{abstract}

\submitto{\JOA}

\maketitle

\section{Introduction}

An interesting class of isotropic inhomogeneous optical media that are currently investigated in optics is represented by the so-called spherical media, i.e. the media with spherically symmetric distribution of refractive index. These media found application in theoretical proposals of many optical devices that provide remarkable optical properties within the framework of geometrical optics. For example, we mention the spherical gradient-index lenses embedded into an optically homogeneous region that provide perfect imaging of two concentric spherical surfaces. Their well known representatives are a Luneburg and an Eaton lens \cite{Luneburg1964,Eaton1952}. In addition, we mention the absolute instruments that provide perfect imaging of three-dimensional domains \cite{BornWolf2006}. Their famous representative is a Maxwell's fish-eye \cite{Maxwell's1854} which, in fact, provides perfect imaging also in terms of wave optics \cite{Leonhardt2009}. 

A common approach to designing the spherical medium of specific properties is based on solving an inverse scattering problem. This is considerably simplified due to the symmetry of the spherical media which enables to treat these media as two-dimensional, because each light ray lies in a plane through the center of symmetry. The classical methods for solving the inverse scattering problem have been derived for both the spherical gradient-index lenses \cite{Luneburg1964} as well as the absolute instruments \cite{Tyc2011a}. Recently, an alternative method based on ideas of transformation optics has also been described \cite{Sarbort2012}. This alternative method utilizes an equivalence of the central section of the spherical medium and a curved rotationally symmetric surface with constant refractive index referred to as a geodesic lens \cite{Rinehart1948,Kunz1954,Cornbleet1981,Sochacki1986}. A great advantage of this method lies in a fact that the shape of a geodesic lens can be deduced relatively easily from the desired light ray trajectories for both the focusing spherical gradient-index lenses and the spherical absolute instruments. Then, once the shape of a geodesic lens is known, the refractive index of an equivalent spherical medium can be calculated by a straightforward procedure. 

In this paper we use this alternative approach based on the concept of geodesic lenses to design new optical devices referred to as multi-focal gradient-index lenses and multi-focal absolute instruments. These devices formed by isotropic spherical media with complex spatial structure provide different imaging properties for different light rays and, therefore, enable to achieve remarkable optical effects. In fact, we have already presented some multi-focal absolute instruments before \cite{Tyc2011a}. However, now we show that the multi-focal absolute instruments as well as the multi-focal gradient-index lenses can easily be designed using the concept of geodesic lenses and that there is a close connection between them. As in our previous paper \cite{Sarbort2012}, we perform the description within the framework of geometrical optics and we utilize the symmetry of spherical media to reduce the description to two dimensions. 

This paper is organized as follows. Section \ref{s_gl_lenses} is devoted to the derivation of the multi-focal gradient-index lenses embedded into an optically homogeneous region that provide perfect imaging of multiple pairs of concentric spherical surfaces. Section \ref{s_AI} is aimed to the derivation of the multi-focal absolute instruments that provide different imaging properties for different spatial domains. Finally, section \ref{Conclusion} concludes the article.

\section{Gradient-index lenses}
\label{s_gl_lenses}

This section is devoted to the spherical gradient-index lenses embedded into an optically homogeneous region. We start with a brief summary of physics related to these optical devices and a procedure for solving the corresponding inverse scattering problem. Then we utilize the general results to design the multi-focal gradient-index lenses that provide perfect imaging of multiple pairs of concentric spherical surfaces.

\subsection{Gradient-index lens and equivalent geodesic lens}

Let us start with a brief summary of physics related to the spherical gradient-index lenses and equivalent geodesic lenses. 

\begin{figure}[ht]
\centering
\includegraphics{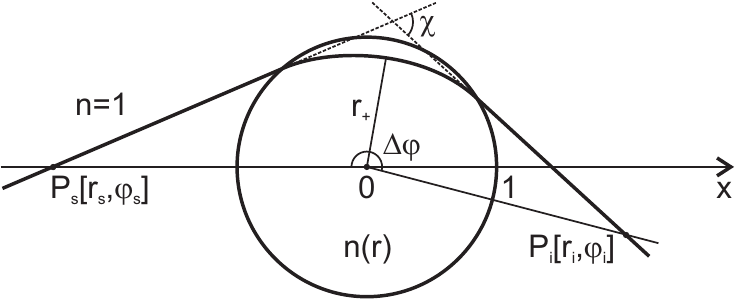}
\caption{Geometrical configuration of the gradient-index lens embedded into an optically homogeneous region.}
\label{f_gil}
\end{figure}

\begin{figure}[ht]
\centering
\includegraphics{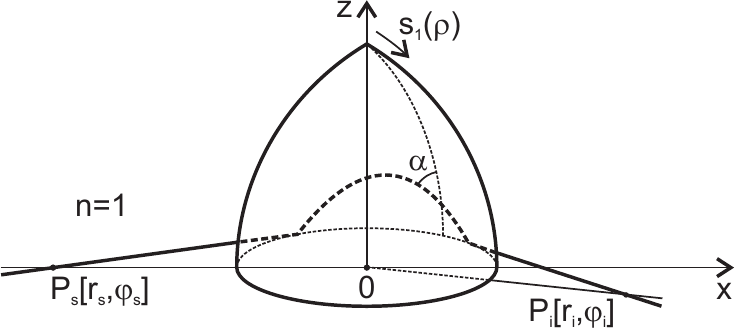}
\caption{Geometrical configuration of the geodesic lens equivalent to the gradient-index lens.}
\label{f_gil_geodesic}
\end{figure}

Consider a plane with the system of polar coordinates $(r, \varphi)$ that represents an optically homogeneous region with refractive index set to unity. Into this plane we place a gradient-index lens of unit radius with the center at the origin (see figure \ref{f_gil}) that is specified by a spherically symmetric refractive index $n(r)$, which meets the boundary condition $n(1)=1$. This gradient-index lens is, in terms of the transformation optics, equivalent to a geodesic lens which is a curved two-dimensional surface (waveguide) with rotational symmetry and refractive index set to unity (see figure \ref{f_gil_geodesic}). We describe its shape by a radial coordinate $\rho$, an angular coordinate $\theta$ and a value of function $s(\rho)$, which represents the length of the surface measured along the meridian from the axis of symmetry to the given point. The coordinate transformation between the gradient-index lens and the geodesic lens follows from the comparison of equivalent optical path elements \cite{Sarbort2012}. Assuming $\theta = \varphi$, we get 
\begin{equation}
\rho = n r, \qquad \mathrm{d}s = n \, \mathrm{d}r .
\label{map}
\end{equation}
Clearly, the shape of a geodesic lens depends primarily on the form of function $\rho(r)$. Within the section \ref{s_gl_lenses} we will assume that for $r \in [0,1]$ the function $ \rho(r)$ is increasing with a single maximum at $r=1$ and, therefore, the shape of a geodesic lens is of the form sketched in figure \ref{f_gil_geodesic}. Although the function $s(\rho)$ is unambiguous, it will be referred to as $s_1(\rho)$ to maintain consistency with section \ref{s_AI}.

Each light ray propagating in the spherical medium or on the geodesic lens is specified by a constant quantity $L$ that is analogous to an angular momentum known from classical mechanics (therefore, we will call the quantity $L$ simply angular momentum). It is given by 
\begin{equation}
L = n r  \sin \alpha = \rho \sin \alpha 
\label{angularmomentum}
\end{equation}
where $\alpha$ is an angle between the tangent to the ray trajectory and the radius vector in the spherical medium or, equivalently, an angle between the tangent to the ray trajectory and the meridian on a geodesic lens. The point of the ray trajectory where $\alpha = \pi/2$ will be referred to as a turning point, its radial coordinate in the spherical medium as $r_+(L)$. Considering an element of the ray trajectory and denoting $s_1'(\rho) = \mathrm{d}s_1(\rho) / \mathrm{d}\rho$, we obtain with help of (\ref{map}) and (\ref{angularmomentum}) a differential equation 
\begin{equation}
\mathrm{d}\varphi = \pm \frac{L\, \mathrm{d}r}{r \sqrt{n^2 r^2 - L^2}} = \pm \frac{L s_1'(\rho)\, \mathrm{d}\rho}{\rho \sqrt{\rho^2 - L^2}} .
\label{dphi}
\end{equation}
that governs propagation of the light ray with angular momentum $L$ in the spherical medium and on the geodesic lens, respectively. These formulas  represent the necessary basis for mathematical formulation of an inverse scattering problem discussed in the following section.

\subsection{Inverse scattering problem}
\label{s_lp_inverse_scattering}

The classical formulation of an inverse scattering problem for a gradient-index lens embedded into an optically homogeneous region lies in derivation of the refractive index $n(r)$ from a given dependence of a scattering angle $\chi(L)$ on the angular momentum $L$, where the scattering angle corresponds to the angle by which the incoming light ray is deflected from its original direction (see figure \ref{f_gil}). Here, we specify bending of the light rays in an equivalent, but slightly different way that is more suitable for addressing the inverse scattering problem related to designing the optical devices that provide stigmatic imaging of given surfaces. 
For a given light ray with angular momentum $L$, we choose a 'source' point P$_{\mathrm{s}}[r_{\mathrm{s}}(L),\varphi_{\mathrm{s}}(L)]$ on the straight line along which the  light ray comes from infinity and we choose an 'image' point P$_{\mathrm{i}}[r_{\mathrm{i}}(L),\varphi_{\mathrm{i}}(L)]$ on the straight line along which the  light ray continues to infinity after passing through the lens. In addition, we introduce an angle $\Delta \varphi(L)$ which corresponds to an absolute value of the polar angle swept by the light ray during its propagation between the points P$_{\mathrm{s}}$ and P$_{\mathrm{i}}$ (see figure \ref{f_gil}). The notation suggests that the radial coordinates $r_{\mathrm{s}}(L)$, $r_{\mathrm{i}}(L)$ as well as the angle $\Delta \varphi(L)$ are, in general, functions of $L$;  this is necessary for considering the multi-focal lenses discussed later. These three functions fully determine bending of the light rays passing through the lens and, therefore, the inverse problem lies in derivation of the refractive index $n(r)$ from the given functions $r_{\mathrm{s}}(L)$, $r_{\mathrm{i}}(L)$ and  $\Delta \varphi(L)$. 

The mathematical formulation of this problem is based on expressing the polar angle $\Delta \varphi(L)$ and it results in an integral equation for the refractive index $n(r)$ of the gradient-index lens or a similar integral equation for the function $s_1'(\rho)$ that parameterizes a geodesic lens. The particular form of these equations depends on the form of function $\rho(r) = n r $. Under the assumptions for the function $\rho(r)$ given above, these integral equations can be written in a compact form 
\begin{equation}
\int_{r_+(L)}^{1} \frac{L\, \mathrm{d}r}{r \sqrt{n^2 r^2 - L^2}} =  \int_{L}^{1} \frac{L\, s_1'(\rho) \mathrm{d}\rho}{\rho \sqrt{\rho^2 - L^2}} = g(L),
\label{lp_index_s_integralequation}
\end{equation}
where the function $g(L)$ is given by
\begin{equation}
g(L) = \frac{1}{2} \left ( \Delta \varphi(L) + \arcsin \frac{L}{r_{\mathrm{s}}(L)} + \arcsin \frac{L}{r_{\mathrm{i}}(L)} - 2 \arcsin L \right ).
\end{equation}
The standard procedure for solving the integral equation for refractive index $n(r)$ can be found in \cite{Luneburg1964}. Here, we use an alternative approach that lies in solving the integral equation for the function $s_1'(\rho)$ and subsequent calculation of the refractive index $n(r)$. 

The integral equation for the function $s_1'(\rho)$ is a kind of Abel integral equation and can be solved by a slight modification of the method presented in \cite{Tyc2011a}. The general result that is independent of the particular form of function $g(L)$ is
\begin{equation}
s_1'(\rho) = - \frac{ 2 \rho}{\pi} \frac{\mathrm{d}}{\mathrm{d} \rho} \left ( \int_{\rho}^{1} \frac{g(L) \mathrm{d}L}{\sqrt{L^2 - \rho^2}} \right ).
\label{ds1drho_general}
\end{equation}
Once we known the function $s_1'(\rho)$, we are able to calculate the refractive index $n(r)$. Using the relations (\ref{map}) we get 
\begin{equation}
\frac {\mathrm{d}r}{r} = s_1'(\rho) \frac {\mathrm{d}\rho}{\rho}.
\label{index_general_start}
\end{equation}
Integrating this equation from the boundary of the lens to the turning point, given in the spherical medium by the radial coordinate $r$ equal to $r_+(L)$ and on the geodesic lens by $\rho$ equal to the angular momentum $L$, we obtain 
\begin{equation}
r_+ (L) = \exp \left ( \int_1^L s_1'(\rho) \frac {\mathrm{d}\rho}{\rho} \right ) . 
\label{rL_exps1}
\end{equation}
Since we have assumed that the function $\rho(r)$ is monotonic, also the function $r_+ (L)$ is monotonic and invertible. Therefore, in principle we can find the inverse function $L(r_+)$ and use it for calculation of the refractive index by means of relation 
\begin{equation}
n(r_+) = \frac{L(r_+)}{r_+} 
\label{lp_nr_general}
\end{equation}
which holds for the turning point because there $\rho = L$. Omitting the lower index of the radial coordinate, we obtain the final result for the refractive index $n(r)$ of the spherical medium forming the gradient-index lens.

\subsection{Multi-focal gradient-index lenses}
\label{s_LP_multifocal_lenses}

In the following we will describe the gradient-index lenses that provide stigmatic imaging of pairs of concentric spherical surfaces. 

In our previous paper \cite{Sarbort2012}, we have discussed the simplest case when there is only one pair of these surfaces. This corresponds to the situation when all the light rays with $L \in [0,1]$ that emerge from a point source located at a given surface and that pass through the lens are focused into an image point located at the second given surface. Such lenses can be found by solving the Luneburg inverse problem \cite{Luneburg1964} and their well-known representatives are the Maxwell's fish-eye lens, the Luneburg lens and the Eaton lens. 

Here, we deal with the multi-focal gradient-index lenses that provide perfect imaging of multiple pairs of concentric spherical surfaces. For simplicity, we start with the case when there are two pairs of these surfaces, i.e. the light rays passing through the lens are divided into two bundles specified by subintervals of angular momentum $[0,L_1]$ and $[L_1,1]$ which, in general, have the source and the image points located at the spherical surfaces of different radii. Then the radial coordinates $r_{\mathrm{s}} (L)$ and $r_{\mathrm{i}} (L)$ of the source and the image points, respectively, can be written as 
\begin{equation}
  r_{\mathrm{s}} (L) = \left\{
  \begin{array}{l}
    r_{\mathrm{s}1} \\
    r_{\mathrm{s}2} \\
  \end{array} \right. ,
  \quad
    r_{\mathrm{i}} (L) = \left\{
  \begin{array}{l l}
    r_{\mathrm{i}1} & \quad \mathrm{for} \quad L \in [0 , L_1]\\
    r_{\mathrm{i}2} & \quad \mathrm{for} \quad L \in (L_1 , 1],\\
  \end{array} \right.
\end{equation}
where $r_{\mathrm{s}1}$, $r_{\mathrm{s}2}$, $r_{\mathrm{i}1}$ and $r_{\mathrm{i}2}$ are given constants greater than or equal to unity. Similarly, we write the angle $\Delta \varphi (L)$ swept by the light rays during the propagation from the source to the image in the form
\begin{equation}
  \Delta \varphi (L) = \left\{
  \begin{array}{l l}
    M_1 \pi & \quad \mathrm{for} \quad L \in [0 , L_1] \\
    M_2 \pi & \quad \mathrm{for} \quad L \in (L_1 , 1] , \\
  \end{array} \right.
\end{equation}
where $M_1 \geq M_2$ are positive constants (a more detailed analysis shows that for $M_1 < M_2$ the integral equation (\ref{lp_index_s_integralequation}) has no solution). The different imaging properties of the light rays corresponding to two subintervals of $L$ imply that the refractive index of the lens must have different functional dependence in two different regions separated by a boundary of radius $r_+(L_1)$. Therefore, we denote
\begin{equation}
  n (r) = \left\{
  \begin{array}{l l}
    n_{11}(r) & \quad \mathrm{for} \quad r \in [0 , r_+(L_1)]\\
    n_{12}(r) & \quad \mathrm{for} \quad r \in [r_+(L_1) , 1] . \\
  \end{array} \right.
\label{LP_index_multifocal}
\end{equation}
Similarly, a geodesic lens equivalent to the gradient-index lens must be parameterized gradually from the top downwards by two different parts of the function $s_1(\rho)$ formally written as 
\begin{equation}
  s_1 (\rho) = \left\{
  \begin{array}{l l}
    s_{11}(\rho) & \quad \mathrm{for} \quad \rho \in [0 , L_1]\\
    s_{12}(\rho) & \quad \mathrm{for} \quad \rho \in [L_1 , 1] . \\
  \end{array} \right.
\label{LP_s1_multifocal}  
\end{equation}

Equipped with the introduced notation, we proceed to solving the inverse scattering problem. Using the general formula (\ref{ds1drho_general}) we get the results for the functions $s_{11}'(\rho)$ and $s_{12}'(\rho)$ in a compact form 
\begin{eqnarray}
s_{11}'(\rho) &=& A_1 + A_2 + \frac{B_1}{\sqrt{1- (\rho/L_1)^2}} + \frac{B_2}{\sqrt{1-\rho^2}}  \nonumber \\
s_{12}'(\rho) &=& A_2 + \frac{B_2}{\sqrt{1-\rho^2}} 
\label{ds11_12drho_multifocal_result} 
\end{eqnarray}
where 
\begin{eqnarray}
A_1 = & \frac{1}{\pi} \left (
\arcsin \sqrt{ \frac{L_1^2-\rho^2}{r_{\mathrm{s}2}^2 - \rho^2}}
- \arcsin \sqrt{ \frac{L_1^2-\rho^2}{r_{\mathrm{s}1}^2 - \rho^2}}  \right. \nonumber \\ 
& + \left. \arcsin \sqrt{ \frac{L_1^2-\rho^2}{r_{\mathrm{i}2}^2 - \rho^2}} 
- \arcsin \sqrt{ \frac{L_1^2-\rho^2}{r_{\mathrm{i}1}^2 - \rho^2}} 
\right ) \nonumber \\
B_1 = & (M_1 - M_2) + \frac{1}{\pi} \left ( \arcsin \frac{L_1}{r_{\mathrm{s}1}}
- \arcsin \frac{L_1}{r_{\mathrm{s}2}} \right. \nonumber \\
& + \left. \arcsin \frac{L_1}{r_{\mathrm{i}1}} - \arcsin \frac{L_1}{r_{\mathrm{i}2}}   
\right ) \nonumber \\
A_2 = & 1 - \frac{1}{\pi} \left ( \arcsin \sqrt{ \frac{1-\rho^2}{r_{\mathrm{s}2}^2 - \rho^2}} + \arcsin \sqrt{ \frac{1-\rho^2}{r_{\mathrm{i}2}^2 - \rho^2}} \right ) \nonumber \\
B_2 = & (M_2 - 1) + \frac{1}{\pi} \left ( \arcsin \frac{1}{r_{\mathrm{s}2}} + \arcsin \frac{1}{r_{\mathrm{i}2}} \right ) .
\label{AB12}
\end{eqnarray}
Clearly, the results for the function $s_{12}'(\rho)$ and the parameters $A_2$ and $B_2$ are formally identical to the results presented in \cite{Sarbort2012} as a solution of the Luneburg inverse problem. 

The refractive index $n(r)$ of the gradient-index lens can be calculated by a general procedure described in section \ref{s_lp_inverse_scattering}. However, an analytical result can be obtained only for the special cases when the radial coordinates of the source and the image points are delimited to only two possibilities -- unity and infinity, and also assuming that $r_{\mathrm{s}1} = r_{\mathrm{s}2}$ and $r_{\mathrm{i}1} = r_{\mathrm{i}2}$. Then the parameter $A_1 = 0$, the parameter $A_2$ becomes independent of $\rho$ and we obtain an implicit formula
\begin{equation}
r^{2/B_2} - 2 r^{1/B_2} (n_{12}r)^{A_2/B_2 - 1} + (n_{12}r)^{2A_2/B_2} = 0 
\label{nr_multifocal}
\end{equation}
for the function $n_{12}(r)$ and a similar formula
\begin{equation}
r = \rho^{A_2 - B_1 - B_2}  \left ( L_1-\sqrt{L_1^2 - \rho^2} \right )^{B_1} \left ( 1-\sqrt{1-\rho^2} \right )^{B_2} 
\label{lp_multifocal_nr}
\end{equation}
which specifies the function $n_{11}(r) = \rho(r)/r$. However, in most cases the last equation can be solved only numerically. 

The described solution of the inverse scattering problem for the gradient-index lens that provide perfect imaging of two pairs of concentric spherical surfaces can easily be generalized for the cases with multiple pairs of these surfaces. Then the light rays passing through the lens are divided into $N$ bundles specified by $N$ subintervals of angular momentum that are separated by $N-1$ real constants $0< L_1 < L_2 < \dots < L_{N-1} < 1$. In the $j$-th subinterval of angular momentum, the functions $r_{\mathrm{s}}(L)$, $r_{\mathrm{i}}(L)$ and $\varphi(L)$ are given by the quantities $r_{\mathrm{s}j}$, $r_{\mathrm{i}j}$ and $M_j \pi$, respectively, which represent generalization of the above notation. The refractive index $n(r)$ is specified by the function $n_{1j}(r)$ within an annular region $r \in [ r_+(L_{j-1}) , r_+(L_{j})]$ and the corresponding part of an equivalent geodesic lens is parameterized by the function $s_{1j}(\rho)$ or by its derivative $s_{1j}'(\rho)$. Using the general formula (\ref{ds1drho_general}), we find the function $s_{1j}'(\rho)$ in a compact form 
\begin{eqnarray}
s_{1j}'(\rho) & = \sum_{k = j}^{N} \left ( A_k + \frac{B_k}{\sqrt{1- ( \rho/L_k )^2}} \right ) 
\label{s1j}
\end{eqnarray}
while we define $L_N = 1$ and the parameters $A_k$ and $B_k$ are given as follows. For $k \in \{ 1,2,\dots N-1 \}$, the parameters $A_k$ and $B_k$ can be obtained from $A_1$ and $B_1$ listed in (\ref{AB12}) by substituting of all indices $1$ with $k$ and $2$ with $k+1$. Similarly, for $k = N$ we get the parameters $A_N$ and $B_N$ from $A_2$ and $B_2$ listed in (\ref{AB12}) by substituting of all indices $2$ with $N$. Finally, once the shape of a geodesic lens is known, the refractive index $n(r)$ can be calculated by the procedure described in section \ref{s_lp_inverse_scattering}.

The general formula (\ref{s1j}) for the function $s_{1j}'(\rho)$ together with the formulas for the parameters $A_k$ and $B_k$ obtained by generalization of (\ref{AB12}) represent the main result of this section. It proves that the approach based on the concept of geodesic lenses is very efficient for designing the multi-focal gradient-index lenses -- in any general case the shape of a geodesic lens can easily be deduced from the desired imaging properties and the refractive index of the gradient-index lens can consequently be calculated by a straightforward procedure.

\subsection{Examples}

\begin{figure}[th]
\centering
\subfloat []{\includegraphics[scale=1.0]{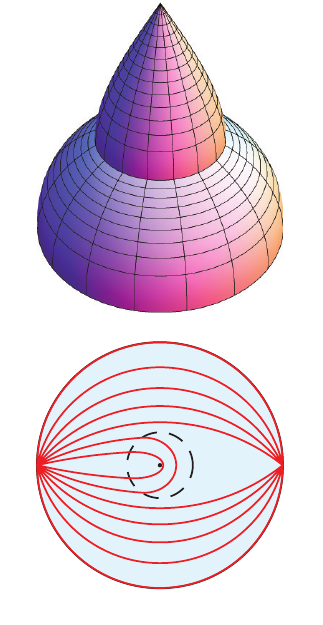} }
\subfloat []{\includegraphics[scale=1.0]{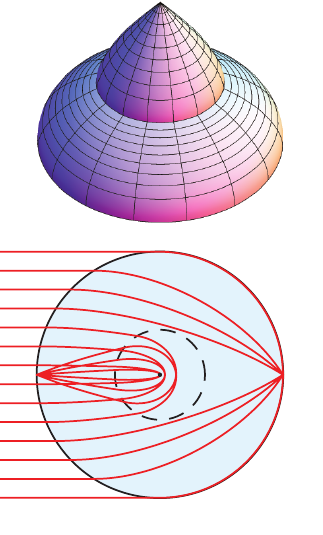} }
\subfloat []{\includegraphics[scale=1.0]{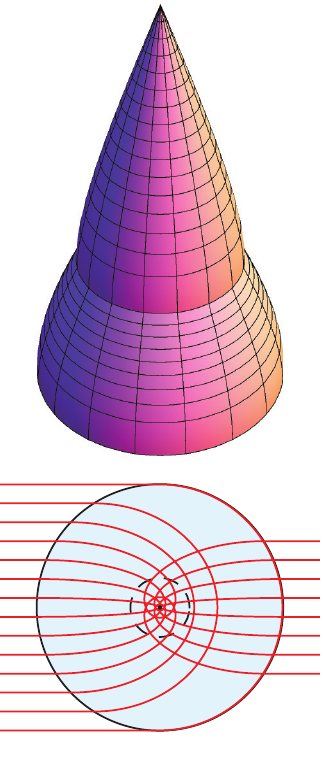} }
\subfloat []{\includegraphics[scale=1.0]{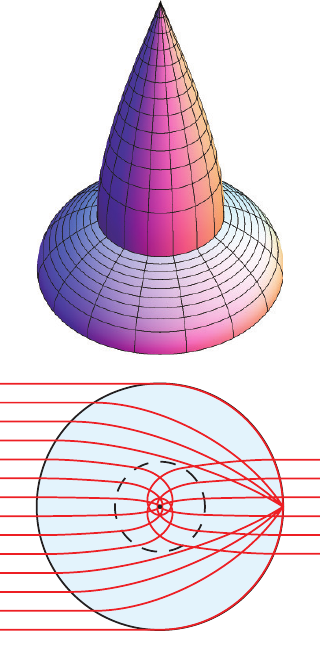} }
\caption{Ray tracing and equivalent geodesic lenses for several multi-focal gradient-index lenses that effectively act as the combination of (a) a classical and a generalized Maxwell's fish-eye, (b) a classical and a generalized Luneburg lens (c) an Eaton lens and an invisible sphere (d) a Luneburg lens and an invisible sphere. The spherical medium is shown in light blue, the dashed circle has the radius $r_+(L_1)$.}
\label{f_multifocal}
\end{figure}

Let us now present in figure \ref{f_multifocal} the ray tracing and the equivalent geodesic lenses for several examples of the multi-focal gradient-index lenses that provide perfect imaging of two concentric spherical surfaces. We focus on the special cases when the source and the image points are located at infinity or on the unit circle and we refer to the lenses according to the effective behavior of the light rays. 

As the first example we present a lens determined by parameters $A_1 = A_2 = 0$, $B_1 = B_2 = 1$ which consists of a classical Maxwell's fish-eye in the outer region, but acts as a generalized Maxwell's fish-eye for the light rays passing through the inner region. The second example is a lens given by numbers $A_1 = 0$, $A_2 = B_2 = \frac{1}{2}$, $B_1 = 1$ corresponding to the combination of a classical and a so-called generalized Luneburg lens which focuses the light rays coming from infinity to two opposite points on the boundary of the lens. As the third example we present a lens with parameters $A_1 = 0$, $A_2 = B_1 = B_2 = 1$ which acts as the combination of an Eaton lens and an invisible sphere. 

\begin{figure}[ht]
\centering
\includegraphics[width=7.5cm]{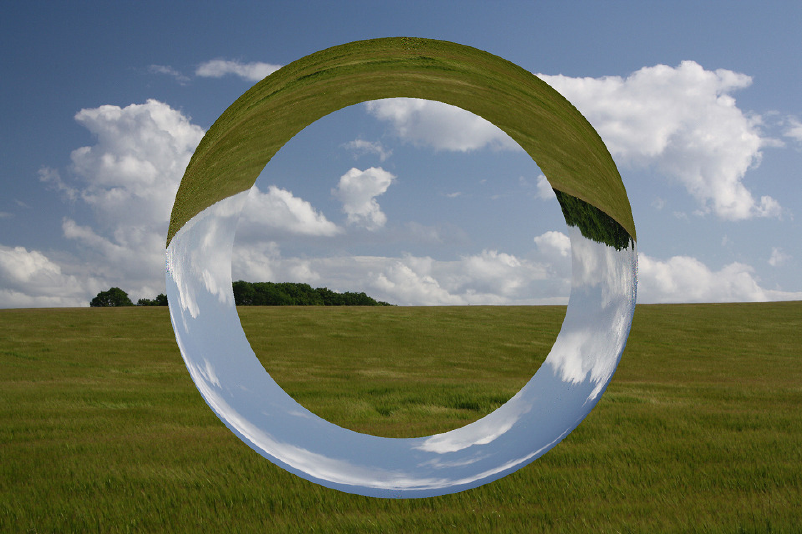}
\caption{A visualization of the optical effect provided by the lens shown in figure \ref{f_multifocal}(d) that effectively acts as the combination of a Luneburg lens and an invisible sphere.}
\label{f_gil_visualization}
\end{figure}

The last example shown in figure \ref{f_multifocal} is probably the most interesting one. This lens effectively acts as the combination of a Luneburg lens and an invisible sphere and, therefore, provides an interesting optical effect. When looking through the lens, the central region shows an image of the scene behind while the peripheral region shows a deformed image of the surrounding scene. This effect is apparent from the visualization shown in figure \ref{f_gil_visualization}.

\section{Absolute instruments}
\label{s_AI}

Another interesting class of optical devices formed by the spherical media are commonly known as absolute instruments. They are characterized by providing perfect imaging of three-dimensional domains within the framework of geometrical optics which can be achieved when all the light rays propagate along closed trajectories. In this section, we briefly review the physics of the spherical media with spatially confined light rays and we solve a corresponding inverse scattering problem. Then we use the general results to design the so-called multi-focal absolute instruments that provide different imaging properties for the light rays with different angular momentum.

\subsection{Spherical medium with spatially confined light rays}

\begin{figure}[ht]
\centering
\includegraphics{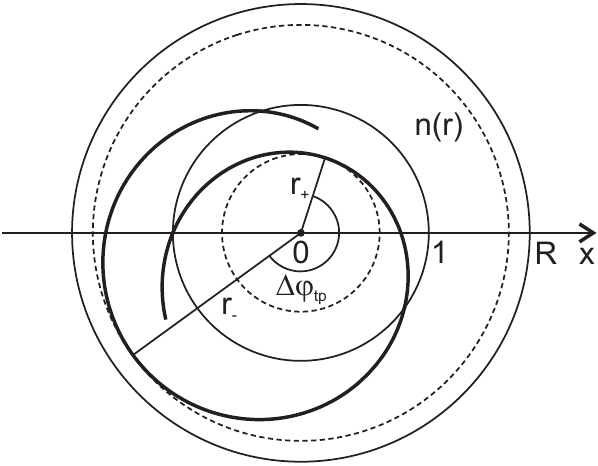}
\caption{Geometrical configuration of the spherical medium with spatially confined light rays.}
\label{f_ai}
\end{figure}

Consider a plane with the system of polar coordinates $(r,\varphi)$ and a spherical medium of radius $R>1$ with the center at the origin that is specified by refractive index $n(r)$. To ensure that all the light rays are spatially confined within this medium (see figure \ref{f_ai}) we assume that the corresponding function $\rho(r) = nr$ has a single global maximum $\rho(1)=1$, the value $\rho(0)=0$ and $\lim_{r \rightarrow R} \rho(r) = 0$. Clearly, the inverse function $r(\rho)$ is multivalued; we denote the corresponding branches by $r_{\pm}(\rho)$. 

\begin{figure}[ht]
\centering
\includegraphics{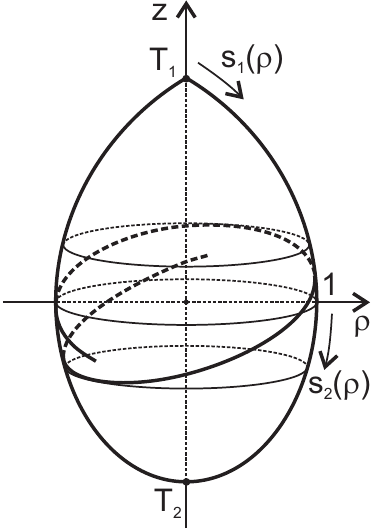}
\caption{Geometrical configuration of the geodesic lens equivalent to the spherical medium with spatially confined light rays.}
\label{f_ai_geodesic}
\end{figure}

The spherical medium defined above is equivalent to a geodesic lens of the shape sketched in figure \ref{f_ai_geodesic}.  Regarding the assumed form of function $\rho(r)$, the geodesic lens must be parameterized by two branches of the function $s(\rho)$ as follows. The branch $s_1(\rho)$ parameterizes the shape of its upper part gradually from the point T$_1$ to the circle $\rho = 1$ referred to as an equator. The branch $s_2(\rho)$ parameterizes its lower part gradually from the equator to the point T$_2$, while we assume $s_2(1) = s_1(1)$. To maintain consistency with section \ref{s_gl_lenses} we assume that the upper and lower part of geodesic lens correspond in the spherical medium to the region inside and outside the unit circle, respectively.

A light ray with angular momentum $L$ propagates in this spherical medium within an annular region bounded by the circles $r_+(L)$ and $r_-(L)$ on which the turning points are located. Equivalently, it propagates on the geodesic lens within a region bounded by circles of radius $\rho = L$ one above and one below the equator on which the turning points are situated. We denote an angle between two consecutive turning points by $\Delta\varphi_{\mathrm{tp}} (L)$. As being justified in \cite{Sarbort2012}, this angle remains the same for infinitely many geodesic lenses that can be mutually reshaped into each other and that are parameterized by functions 
\begin{eqnarray}
s_1'(\rho) &=& s_{\mathrm{a}}'(\rho) + s_{\mathrm{s}}'(\rho) \nonumber  \\
s_2'(\rho) &=& s_{\mathrm{a}}'(\rho) - s_{\mathrm{s}}'(\rho) 
\label{s1s2der_sass}
\end{eqnarray}
where the functions $s_{\mathrm{s}}'(\rho)$ and $s_{\mathrm{a}}'(\rho)$ describe the symmetric and the antisymmetric part of a geodesic lens, respectively. This represents a key fact that allows  solving an inverse scattering problem discussed in the following section.

\subsection{Inverse scattering problem}

An inverse scattering problem for spatially confined light rays lies in derivation of the refractive index $n(r)$ from the given function $\Delta\varphi_{\mathrm{tp}} (L)$. Expressing this angle by means of equation (\ref{dphi}), we get integral equations 
\begin{equation}
\int_{r_-(L)}^{r_+(L)}\frac{L\, \mathrm{d}r}{r \sqrt{n^2 r^2 - L^2}} 
= \int_{L}^{1}\frac{L\, ( s_1'(\rho) - s_2'(\rho) ) \mathrm{d}\rho}{\rho \sqrt{\rho^2 - L^2}}
=  \Delta\varphi_{\mathrm{tp}} (L) 
\label{ai_int_eq}
\end{equation}
for the refractive index $n(r)$ and the functions $s_1'(\rho)$ and $s_2'(\rho)$ that parameterize the geodesic lens. The procedure for solving the integral equation for the refractive index $n(r)$ can be found in \cite{Tyc2011a}. Here, we use an alternative approach already used in section \ref{s_gl_lenses} that lies in solving the integral equation for the functions $s_1'(\rho)$ and $s_2'(\rho)$ and subsequent calculation of the refractive index $n(r)$. 

The procedure for solving the integral equation (\ref{ai_int_eq}) starts by employing the relations (\ref{s1s2der_sass}). Then the integral equation is reduced to 
\begin{eqnarray}
\int_{L}^{1} \frac{L\, s_{\mathrm{s}}'(\rho) \mathrm{d}\rho}{\rho \sqrt{\rho^2 - L^2}} = \frac{1}{2} \Delta\varphi_{\mathrm{tp}} (L)
\label{ai_integralequationgeodesic_g}
\end{eqnarray}
where the only unknown function is $s_{\mathrm{s}}'(\rho)$. This is again a kind of Abel integral equation and its general solution is 
\begin{equation}
s_{\mathrm{s}}'(\rho) = - \frac{  \rho}{\pi} \frac{\mathrm{d}}{\mathrm{d} \rho} \left ( \int_{\rho}^{1} \frac{ \Delta\varphi_{\mathrm{tp}} (L) \mathrm{d}L}{\sqrt{L^2 - \rho^2}} \right ).
\label{ai_dssdrho_general}
\end{equation}
This formula allows to calculate the shape of a symmetrical geodesic lens from a given function  $\Delta\varphi_{\mathrm{tp}} (L)$. Then, by choosing the function $s_{\mathrm{a}}'(\rho)$ that makes the geodesic lens asymmetrical, it is possible to design a variety of geodesic lenses that represent spherical media with a common function $\Delta\varphi_{\mathrm{tp}} (L)$. 

The refractive index $n(r)$ of the spherical medium can be calculated by means of the method described in section \ref{s_lp_inverse_scattering}. In fact, the described method is directly applicable to derive the refractive index within the unit circle. The refractive index outside the unit circle can be calculated by formally identical relations obtained by a simple substitution $s_1'(\rho) \rightarrow s_2'(\rho)$ and $r_+(L) \rightarrow r_-(L)$.

\subsection{Multi-focal absolute instruments}

Let us now utilize the general results for discussion of the absolute instruments that provide perfect imaging of three-dimensional domains. 

In our previous paper \cite{Sarbort2012}, we have discussed the simplest case when the angle $\Delta\varphi_{\mathrm{tp}} (L)$ swept by the light ray during the propagation between two consecutive turning points is independent of $L$ and equals to a rational fraction of $\pi$. Then the ray trajectories are closed and each source point is sharply imaged into a set of image points depending on the character of ray trajectories.  

Here, we deal with the so-called multi-focal absolute instruments that provide different imaging properties for the light rays with different angular momentum. Analogously to section \ref{s_gl_lenses}, we start with the case when the angle $\Delta \varphi_{\mathrm{tp}} (L)$ is piecewise constant on two subintervals of $L$. We denote 
\begin{equation}
  \Delta \varphi_{\mathrm{tp}} (L) = \left\{
  \begin{array}{l l}
    \mathcal{B}_1 \pi & \quad \mathrm{for} \quad L \in [0 , L_1]\\
    \mathcal{B}_2 \pi & \quad \mathrm{for} \quad L \in (L_1 , 1] , \\
  \end{array} \right.
\label{ai_multifocal_deltaphitp}
\end{equation}
where $\mathcal{B}_1 \geq \mathcal{B}_2$ are positive rational numbers and $ 0 < L_1 < 1$ (the condition for $\mathcal{B}_1$ and $\mathcal{B}_2$ is analogous to the condition $M_1 \geq M_2$ mentioned in section \ref{s_LP_multifocal_lenses}). This assumption implies that the refractive index of the spherical medium has different functional dependence within four annular regions separated by the boundaries of radii $r_+(L)$, $1$ and $r_-(L)$, respectively. For the refractive index inside the unit circle we use the notation (\ref{LP_index_multifocal}), in addition we denote 
\begin{equation}
  n (r) = \left\{
  \begin{array}{l l}
    n_{22}(r) & \quad \mathrm{for} \quad r \in [1 , r_-(L_1)]\\
    n_{21}(r) & \quad \mathrm{for} \quad r \in [r_-(L_1) , R] . \\
  \end{array} \right.
\label{AI_index_multifocal}
\end{equation}
An equivalent geodesic lens is parameterized by two different parts of functions $s_1 (\rho)$ and $s_2 (\rho)$. The former is already specified by (\ref{LP_s1_multifocal}), the latter is given by 
\begin{equation}
  s_2 (\rho) = \left\{
  \begin{array}{l l}
    s_{22}(\rho) & \quad \mathrm{for} \quad \rho \in [L_1 , 1]\\
    s_{21}(\rho) & \quad \mathrm{for} \quad \rho \in [0 , L_1], \\
  \end{array} \right.
\end{equation}
where the functions $s_{22}(\rho)$ and $s_{21}(\rho)$ parameterize the lower part of a geodesic lens gradually from the equator downwards. Similarly, we split the functions $s_{\mathrm{s}}(\rho)$ and $s_{\mathrm{a}}(\rho)$ as 

\begin{equation}
  s_{\mathrm{s}}(\rho) = \left\{
  \begin{array}{l}
    s_{\mathrm{s}1}(\rho) \\
    s_{\mathrm{s}2}(\rho) \\
  \end{array} \right. ,
  \quad
    s_{\mathrm{a}}(\rho) = \left\{
  \begin{array}{l l}
    s_{\mathrm{a}1}(\rho) & \quad \mathrm{for} \quad \rho \in [0 , L_1]\\
    s_{\mathrm{a}2}(\rho) & \quad \mathrm{for} \quad \rho \in [L_1 , 1].\\
  \end{array} \right.
\end{equation}

Equipped with the notation, we proceed to solving the inverse scattering problem. Using the general formula (\ref{ai_dssdrho_general}) we get the functions 
\begin{eqnarray}
s_{\mathrm{s}1}'(\rho) = & \frac{\mathcal{B}_1-\mathcal{B}_2}{\sqrt{1- ( \rho/L_1 )^2}} + \frac{\mathcal{B}_2}{\sqrt{1-\rho^2}} \nonumber \\
s_{\mathrm{s}2}'(\rho) = & \frac{\mathcal{B}_2}{\sqrt{1-\rho^2}} 
\label{ai_ss12}
\end{eqnarray}
that describe a geodesic lens symmetrical with respect to the equatorial plane. Choosing the functions $s_{\mathrm{a}1}'(\rho)$ and $s_{\mathrm{a}2}'(\rho)$ we get a variety of asymmetrical geodesic lenses that are equivalent to the multi-focal absolute instruments. 

In analogy with the results of section \ref{s_LP_multifocal_lenses}, we focus on a class of geodesic lenses given by the choice  $s_{\mathrm{a}1}'(\rho) = \mathcal{A}_1$ and $s_{\mathrm{a}2}'(\rho) = \mathcal{A}_2$, where $\mathcal{A}_1$ and $\mathcal{A}_1$ are non-negative real numbers. Then the geodesic lens is described by a set of functions
\begin{eqnarray}
s_{11}'(\rho) =& \mathcal{A}_1 + \frac{\mathcal{B}_1-\mathcal{B}_2}{\sqrt{1- ( \rho/L_1 )^2}} + \frac{\mathcal{B}_2}{\sqrt{1-\rho^2}} \nonumber \\
s_{12}'(\rho) =& \mathcal{A}_2 + \frac{\mathcal{B}_2}{\sqrt{1-\rho^2}} \nonumber \\
s_{22}'(\rho) =& \mathcal{A}_2 - \frac{\mathcal{B}_2}{\sqrt{1-\rho^2}} \nonumber \\
s_{21}'(\rho) =& \mathcal{A}_1 - \frac{\mathcal{B}_1-\mathcal{B}_2}{\sqrt{1- ( \rho/L_1 )^2}} - \frac{\mathcal{B}_2}{\sqrt{1-\rho^2}} .
\label{ai_multifocal_sij}
\end{eqnarray}
In such case, the refractive index of the spherical medium can be calculated by means of equations derived in section \ref{s_LP_multifocal_lenses}. The functions $n_{12}(r)$ and $n_{22}(r)$ are given by a formula obtained from (\ref{nr_multifocal}) by a simple substitution $A_2 \rightarrow \mathcal{A}_2$ and $B_2 \rightarrow \pm \mathcal{B}_2$. Similarly, the functions  $n_{11}(r)$ and $n_{21}(r)$ are given by an equation obtained from (\ref{lp_multifocal_nr}) by a substitution $A_1 \rightarrow \mathcal{A}_1$,  $B_1 \rightarrow \pm (\mathcal{B}_1 - \mathcal{B}_2)$ and $B_2 \rightarrow \pm \mathcal{B}_2$.

Similarly as in section \ref{s_gl_lenses}, the described solution of the inverse scattering problem for the absolute instruments that provide different imaging for the light rays specified by two subintervals of angular momentum can easily be generalized for the cases with $N$ of these subintervals separated by $N-1$ real constants $0< L_1 < L_2 < \dots < L_{N-1} < 1$. Then the function $\Delta\varphi_{\mathrm{tp}} (L)$ is given in the $j$-th subinterval of angular momentum by the quantity $\mathcal{B}_j \pi$, which represents generalization of the above notation. The refractive index is given by the functions $n_{1j}(r)$ and $n_{2j}(r)$ inside and outside the unit circle, respectively. An equivalent geodesic lens is parameterized by the functions $s_{1j}(\rho)$ and $s_{2j}(\rho)$ above and below the equator, respectively, and its symmetric and antisymmetric part are described by functions $s_{\mathrm{s}j}(\rho)$ and $s_{\mathrm{a}j}(\rho)$, respectively. Using the general formula (\ref{ai_dssdrho_general}), we find the function $s'_{\mathrm{s}j}(\rho)$ in a compact form 
\begin{equation}
s_{\mathrm{s}j}'(\rho) = \sum_{k = j}^{N} \frac{\mathcal{B}_k - \mathcal{B}_{k+1}}{\sqrt{1- ( \rho/L_k )^2}} 
\label{s_sj}
\end{equation}
while we define $L_N = 1$ and $B_{N+1} = 0$. Choosing the functions $s_{\mathrm{a}j}'(\rho)$ we get a variety of asymmetrical geodesic lenses and, finally, we calculate the refractive index of the corresponding multi-focal absolute instruments by the procedure described in section \ref{s_lp_inverse_scattering}.

The general formula (\ref{s_sj}) represents the main result of this section. It proves that the approach based on the concept of geodesic lenses is very efficient also for designing the multi-focal absolute instruments. Moreover, similarity of the general formulas (\ref{s1j}) and (\ref{s_sj}) reveals that there is a close connection between the multi-focal gradient-index lenses and the multi-focal absolute instruments.

\subsection{Examples}

\begin{figure}[t]
\centering
\subfloat []{\includegraphics{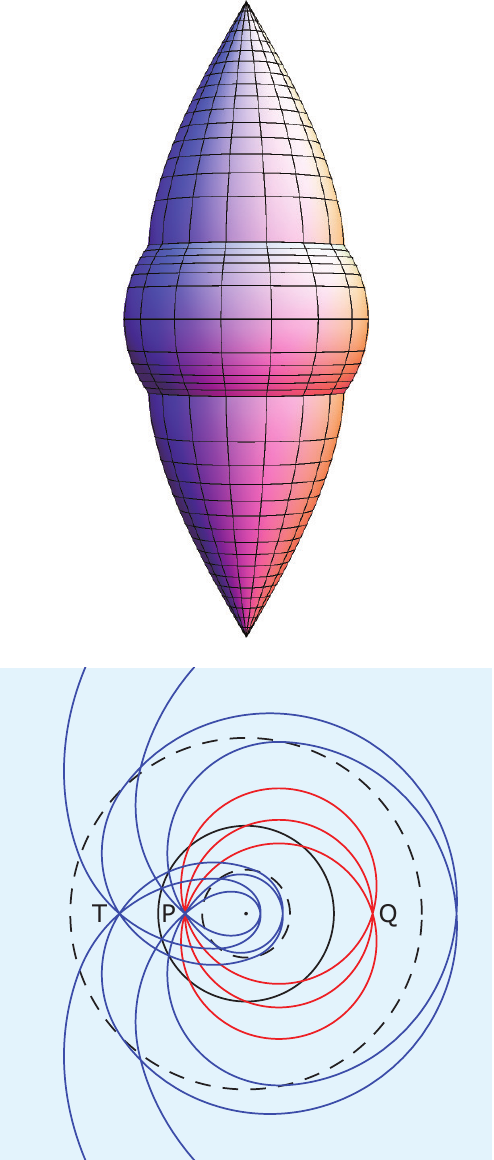} }
\subfloat []{\includegraphics{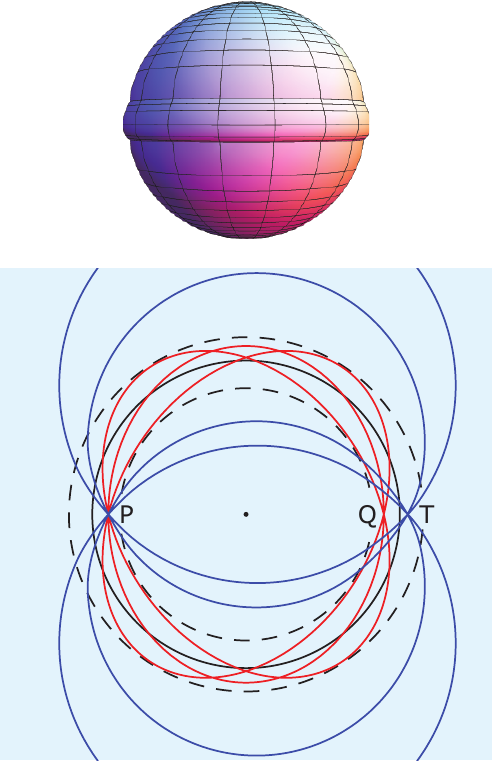} }
\subfloat []{\includegraphics{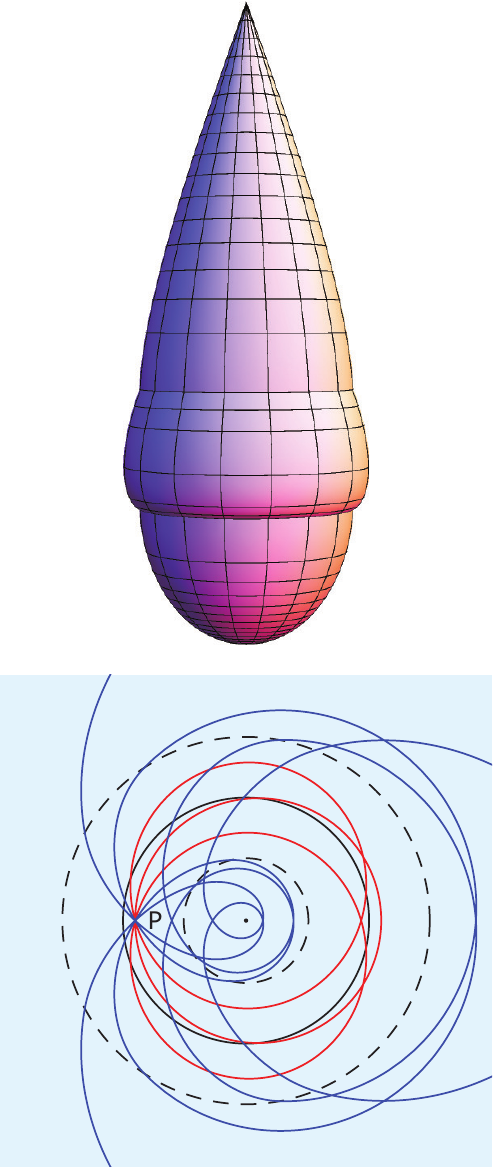} }
\caption{Ray tracing and equivalent geodesic lenses for the multi-focal absolute instruments given by the following parameters: (a) $\mathcal{A}_1 = \mathcal{A}_2 =0$, $\mathcal{B}_1 = 2$, $\mathcal{B}_2 =1$, (b) $\mathcal{A}_1 = 0$, $\mathcal{A}_2 = \frac{1}{2}$, $\mathcal{B}_1 = 1$, $\mathcal{B}_2 = \frac{1}{2}$,  (c)  $\mathcal{A}_1 = \mathcal{A}_2 = \mathcal{B}_2 = 1$, $\mathcal{B}_1 =2$. The black circle corresponds to the circular trajectory at $r = 1$ with maximum possible angular momentum $L = 1$, the dashed circles have the radii $r_{\pm}(L_1)$. }
\label{f_ai_multifocal}
\end{figure}

Let us now show in figure \ref{f_ai_multifocal} the ray tracing and the geodesic lenses corresponding to several examples of the multi-focal absolute instruments that provide different imaging properties for the light rays specified by two subintervals of angular momentum. \

As the first example shown in figure \ref{f_ai_multifocal}(a) we present an absolute instrument specified by parameters $\mathcal{A}_1 = \mathcal{A}_2 =0$, $\mathcal{B}_1 = 2$ and $\mathcal{B}_2 =1$. This corresponds to the simple case when the geodesic lens is symmetrical with respect to the equatorial plane. Its central part corresponding to $\rho \geq L_1$ is formed by a spherical surface on which the light rays propagate along great circles, the parts corresponding to $\rho \leq L_1$ one above and one below the equator are formed by more complex surfaces on which the light rays propagate along the geodesic curves that cannot be simply classified. The refractive index of an equivalent spherical medium is given by the profile of Maxwell's fish-eye within an annular region bounded by the dashed circles of radii $r_{\pm}(L_1)$ and by the numerically calculated profiles outside this annular region. The light rays emerging from a point P are divided into two color-coded bundles corresponding to two subintervals of $L$. The red light rays with $L \in (L_1,1]$ propagate along the circular trajectories within the dashed bounded annulus and they meet at the image point Q before they reach the starting point P again. On the other hand, the blue light rays with $L \in [0,L_1]$ propagate also outside the dashed bounded annulus and they meet at the image point T before returning to the point P. The blue and the red trajectories also differ in the polar angle between two consecutive turning points according to the choice of parameters $\mathcal{B}_1$ and $\mathcal{B}_2$, respectively.
 
The second example shown in figure \ref{f_ai_multifocal}(b) is specified by parameters $\mathcal{A}_1 = 0$, $\mathcal{A}_2 = \frac{1}{2}$, $\mathcal{B}_1 = 1$ and $\mathcal{B}_2 = \frac{1}{2}$. The geodesic lens is asymmetrical with respect to the equatorial plane and the refractive index is given by a Luneburg profile within the dashed bounded annulus. The red light rays with $L \in (L_1,1]$ emerging from a point P propagate along the concentric ellipses and meet at the image point Q before returning to the point P. The blue light rays with $L \in [0,L_1]$ emerging from the point P meet at the image point T that is different from Q and the angle between the consecutive turning points is twice that of the red rays. 

Finally, we show in figure \ref{f_ai_multifocal}(c) the third example given by parameters $\mathcal{A}_1 = \mathcal{A}_2 = \mathcal{B}_2 = 1$ and $\mathcal{B}_1 =2$. In this case, the refractive index is given by an Eaton profile within the dashed bounded annulus and, therefore, the light rays propagate within this region along the confocal ellipses. The red rays with $L \in (L_1,1]$ emerging from a point P return to this point after making one loop around the origin. On the other hand, the blue rays with $L \in [0,L_1]$ emerging from the point P need to make two loops around the origin to close their trajectory.

\section{Conclusion}
\label{Conclusion}

In this paper we have presented a general approach to designing isotropic spherical media with complex spatial structure that provide different types of imaging for different light rays. This approach  based on equivalence of the spherical medium and the corresponding geodesic lens proved to be very efficient for designing the multi-focal gradient-index lenses embedded into an optically homogeneous region as well as the multi-focal absolute instruments that provide perfect imaging of three-dimensional domains.

\ack
TT acknowledges support from grant no. P201/12/G028 of the Grant Agency of the Czech Republic and from the QUEST programme grant of the Engineering and Physical Sciences Research Council.


\section*{References}
\begin{thebibliography}{20}
\bibitem{Luneburg1964} Luneburg R K 1964 {\it Mathematical Theory of Optics\/} (Berkeley: University of California Press)
\bibitem{Eaton1952} Eaton J E 1952 {\it Trans. IRE Antennas Propag.} 4 66 
\bibitem{BornWolf2006} Born M and Wolf E 2006 {\it Principles of optics\/} (Cambridge: Cambridge University Press)
\bibitem{Maxwell's1854} Maxwell's J C 1854 {\it Camb. Dublin Math. J. } 8 188
\bibitem{Leonhardt2009} Leonhardt U 2009 {\it New J. Phys.} 11 093040
\bibitem{Tyc2011a} Tyc T, Herz\'{a}nov\'{a} L, \v{S}arbort M and Bering K 2011 {\it New J. Phys.} 13 033016
\bibitem{Sarbort2012} \v{S}arbort M and Tyc T 2012 {\it J. Opt.} 14 075705
\bibitem{Rinehart1948} Rinehart R F 1948 {\it J. Appl. Phys.} 19 860  
\bibitem{Kunz1954} Kunz K S 1954 {\it J. Appl. Phys.} 25 642
\bibitem{Cornbleet1981} Cornbleet S and Rinous P J 1981 {\it IEE Proc-H} 128 95
\bibitem{Sochacki1986} Sochacki J 1986 {\it Appl. Optics} 25 235
\bibitem{Firsov1953} Firsov O B 1953 {\it Zh. Eksp. Teor. Fiz.} 24 279
\bibitem{Minano2006a} Mi\~{n}ano J C 2006 {\it Opt. Express} 14 9627
\bibitem{Gutman1954} Gutman A S 1954 {\it J. Appl. Phys.} 25 855
\bibitem{Morgan1958} Morgan S P 1958 {\it J. Appl. Phys.} 29 1358
\end{thebibliography}
\end{document}